\documentstyle[12pt,aasms4]{article}

\received{19 February 1996}
\accepted{31 May 1996}

\begin{document}

\title{Environmental Effect on the Associations of Background
      Quasars with Foreground Objects: I. Analytic Investigation}

\author{Xiang-Ping Wu}
\affil{Department of Physics, University of Arizona, Tucson, AZ 85721 and \\
       Beijing Astronomical Observatory, Chinese Academy of Sciences,
       Beijing 100080, China}

\author{Li-Zhi Fang}
\affil{Department of Physics, University of Arizona, Tucson, AZ 85721}

\author{Zonghong Zhu}
\affil{Department of Astronomy, Beijing Normal University, 
       Beijing 100875, China}

\and

\author{Bo Qin}
\affil{Beijing Astronomical Observatory, Chinese Academy of Sciences,
       Beijing 100080, China}

\begin{abstract}
The associations of the angular positions of background quasars
with foreground galaxies, clusters of galaxies and quasars are
often attributed to the statistical lensing by gravitational 
potentials of the matter along the lines of sight, although
it has been known that none of the individual objects (galaxies, 
clusters or quasars) are able to fully explain the reported 
amplitudes of the quasar number enhancements. This probably 
arises from the fact that the gravitational lensing effect by 
the environmental matter surrounding these objects has been 
ignored. In this paper we conduct an extensive study of 
the influence of the environmental matter on the prediction 
of quasar enhancement factor by employing the spatial two-point 
correlation function. Assuming a singular isothermal sphere 
for mass density profile in galaxy and cluster of galaxies, 
we estimate the average surface  mass density 
$\overline{\Sigma}$ around galaxies, clusters 
and quasars from the galaxy-galaxy, cluster-cluster, 
cluster-galaxy and quasar-galaxy correlations.
Our results show that the $\overline{\Sigma}$ induced 
quasar number enhancement in the scenario of gravitational
magnification depends critically on the mass density parameters
of galaxies ($\Omega_g$) and clusters of galaxies ($\Omega_c$)
in the universe. For a flat cosmological model of $\Omega_0=1$ 
the environmental matter can indeed play an important role 
in the lensing origin of the quasar-quasar and quasar-galaxy 
associations if $\Omega_g\sim\Omega_c\sim\Omega_0$, 
while it is unlikely that $\overline{\Sigma}$ is sufficient 
to account for the reported quasar overdensity behind  
quasars/galaxies if galaxies and clusters of galaxies 
contribute  no more than $25\%$ to the matter of the universe. 
Nonetheless, the recently observed quasar-cluster associations on 
scale of $\sim10$ arcminutes cannot be 
the result of gravitational lensing by the cluster environmental 
matter even if $\Omega_g=\Omega_c=\Omega_0$.
\end{abstract}

\keywords{gravitational lensing -- 
          large-scale structure of universe}

\section{Introduction}

Detection of the associations of the angular positions of
background quasars with foreground
galaxies and clusters of galaxies
on scale of $\sim10$ arcminutes in recent years
(Fugmann 1988, 1990; Bartelmann \& Schneider 1993b, 1994; 
Rodrigues-Williams \& Hogan 1994; Wu \& Han 1995; Rodrigues-Williams 
\& Hawkins 1995; Seitz \& Schneider 1995)
has led several studies on the statistical lensing of quasars by 
large-scale structures traced by galaxies and clusters of galaxies
(e.g. Bartelmann \& Schneider 1993a, b; Wu \& Fang 1996). 
Indeed, gravitational matter of galaxies and clusters of galaxies
alone is far from providing the observed amplitude of quasar
overdensity to that large angular scale 
[see Wu (1996) for a recent review]. 
Even on small scale of a few arcseconds, the
quasar-galaxy associations cannot be well accounted for unless 
an unreasonably large velocity dispersion for galaxies or 
an additional matter contribution is assumed
(e.g. Narayan 1989; Schneider 1989; Wu, Zhu \& Fang 1996). 
In particular, the quasar environmental matter must be invoked in the
scenario of gravitational lensing 
for the explanation of the  existence of four 
quasar pairs within $5^{\prime\prime}$ but with different redshifts 
among the 1000 -- 2000 surveyed quasars 
(Burbidge, Hoyle \& Schneider 1996).

The spatial two-point correlation function is a quantitative 
description of the environmental matter distribution.  
Anderson \& Alcock (1986) addressed the question whether clustering of
galaxies described by the two-point correlation function alters the
statistical properties of gravitational lensing. From a number of
Monte-Carlo simulations, they concluded that the
effect is too small to be significant for statistics of 
the multiply-imaged quasars in the case of galaxies acting as lenses.  
However, based on the N-body simulations of formation of galaxies,
Bartelmann \& Schneider (1993a) 
did find a correlation of high redshift quasars with low redshift
galaxies, which stems from the magnification bias by 
galaxies and their surrounding large-scale structures. The 
results from these two studies are not inconsistent since they 
investigated two different phenomena though the deflectors are 
the same (galaxies).

It is necessary to further explore a more general question:
Is gravitational lensing by the environmental 
matter able to explain the reported 
associations of background quasars with foreground objects 
or how large is the environmental effect on the prediction
of the background quasar overdensity around foreground objects ?
We intend to answer the question in this paper 
by an extensive analytic investigation of the 
magnification bias from the two-point correlation functions
and will present elsewhere (Wu, Fang \& Jing, in preparation)
a numerical study of the issue from the
N-body simulations
of formation of clustering of galaxies and clusters of galaxies.
In section 2 we estimate the mean surface matter density 
superposed statistically
on galaxies, clusters of galaxies and quasars from 
the two-point correlation  functions of galaxy-galaxy, cluster-cluster,
cluster-galaxy and quasar-galaxy. We then evaluate the quasar enhancement
factor near the positions of foreground galaxies, clusters and quasars
by the environmental effect in section 3.
A brief discussion is presented in section 4. 
Throughout the paper we adopt a flat cosmological model of
$\Omega_0=1$ and a Hubble constant of $H_0=100\;h$ km s$^{-1}$ Mpc$^{-1}$.

\section{Environmental
matter contribution from the two-point correlation function}

\subsection{General consideration}

The spatial two-point correlation function can be generally written as
\begin{equation}
\xi_{AB}(r)=\left(\frac{r}{r_0}\right)^{-\gamma}
\end{equation}
at the range of separation $r_{min}\leq r\leq r_{max}$, where $r_0$ is
the correlation length. $\xi_{AB}(r)$ provides a conditional probability
of finding neighbor objects $B$ (quasars, galaxies, clusters, etc.) 
in the comoving volume $dV$ at a comoving distance $r$ from a 
given object $A$ (quasar, galaxy, cluster, etc.).   
We concentrate ourselves on the plane perpendicular to the line of sight 
at the position of object $A$ with redshift $z_d$. The expected number 
of objects $B$ within a radius $d\zeta$ of $\zeta$ from $A$ is
\begin{equation}
dN_B(\zeta)=4\pi n_{B0}\zeta d\zeta
           \int_{\zeta}^{r_{max}}[1+\xi_{AB}(r)(1+z_d)^{\epsilon}]
	        \frac{rdr}{\sqrt{r^2-\zeta^2}},
\end{equation}
where $n_{B0}$ is the present number density of objects $B$ and 
$\epsilon$ accounts for the evolution of the correlation function.  
In the following we consider only
those excess population of objects $B$ relative to the ``background''
ones, i.e., we exclude the contribution of the mean number density 
of objects $B$.

If we assume that the mass density profile of  object $B$ has the form 
of a singular isothermal sphere (SIS) 
with velocity dispersion $\sigma_B$,
which is a good approximation for the dark matter distribution in
galaxies and in clusters of galaxies, the mass contribution
from $B$ at position $\zeta$ to an area of $\pi\zeta_0^2$ is simply
\begin{equation}
m_B(\zeta_0,\zeta)=\left\{
\begin{array}{ll}
\frac{2\sigma_B^2}{G}\;\zeta_0\;\int_0^{\sin^{-1}(\zeta_0/\zeta)}
           \;\kappa(\zeta,\theta)\; d\theta, \;\;\; &
           \zeta_0<\zeta;\\
\frac{\sigma_B^2}{G}\;\zeta_0\;\int_0^{\pi}
          \;\kappa(\zeta,\theta)\; d\theta, \;\;\; & 
           \zeta_0>\zeta,
\end{array} \right.
\end{equation}
in which $\kappa(\zeta,\theta)=\sqrt{1-(\zeta/\zeta_0)^2\sin^2\theta}$.
The expected mass contribution to the area $\pi\zeta_0^2$  
from an ensemble of objects $B$   is thus 
\begin{equation}
m(\zeta_0)=\int\;m_B(\zeta_0,\zeta)\;dN_B(\zeta).
\end{equation}
Therefore, we can obtain the expected mean surface mass 
density of objects $B$ around a given object $A$ simply
by $m(\zeta_0)/\pi\zeta_0^2$, which reads 
\begin{equation}
\Sigma_{AB}(\zeta_0)=4 n_{B0}(1+z_d)^{\epsilon} r_0^2 
              \left(\frac{\sigma_B^2}{G}\right)\;\tilde{F}(\zeta_0),
\end{equation}
where
\begin{equation}
\tilde{F}(\zeta_0)=
\left(\frac{\zeta_0}{r_0}\right)^{2-\gamma}
\left[
\int_0^{\zeta_0}\chi(\zeta)d(\zeta/\zeta_0)
\int_0^{\pi}\kappa(\zeta,\theta)d\theta + 
2\int_1^{r_{max}}\chi(\zeta)d(\zeta/\zeta_0)
\int_0^{\sin^{-1}(\zeta_0/\zeta)}\kappa(\zeta,\theta)d\theta
\right],
\end{equation}
and 
\begin{equation}
\chi(\zeta_0,\zeta)=\left\{
\begin{array}{ll}
     (\zeta/\zeta_0)^{2-\gamma}
    \int_{1}^{r_{max}/\zeta}
           \frac{dx}{x^{\gamma-1}\sqrt{x^2-1}}, \;\;\;\; &
                               r_{min}<\zeta\leq r_{max};\\
    (\zeta/\zeta_0)^{2-\gamma} 
    \int_{r_{min}/\zeta}^{r_{max}/\zeta}
           \frac{dx}{x^{\gamma-1}\sqrt{x^2-1}}, \;\;\;\; &
                              \zeta\leq r_{min}.
\end{array}  \right.
\end{equation}

Suppose that the objects $B$ follow a luminosity distribution function
$\phi_B(L_B)dL_B$ and the luminosity $L_B$ is related to the velocity
dispersion through
\begin{equation}
\frac{L_B}{L_B^*}=\left(\frac{\sigma_B}{\sigma_B^*}\right)^{\nu},
\end{equation}
in which $L_B^*$ and $\sigma_B^*$ are the characteristic luminosity and
corresponding velocity dispersion, respectively. The total
surface mass density given by  $\xi_{AB}(r)$ is the sum of 
all the objects $B$ over their distribution of velocity dispersion
or luminosity:
\begin{equation}
\overline{\Sigma}_{AB}(\zeta_0)=4(1+z_d)^{\epsilon} r_0^2 
		\frac{(\sigma_B^*)^2}{G} 
           \tilde{F}(\zeta_0)
          \int_0^{\infty} \left(\frac{L_B}{L_B^*}\right)^{2/\nu}
          \phi_B(L_B)dL_B.
\end{equation}

However, applying SIS model to an arbitrarily large radius
may lead to an overestimate of the matter contribution 
of objects $B$. A reasonable hypothesis is that the SIS profile
is truncated at a radius $R_B$ so that the total mass of object
$B$ is $M_B=2R_B\sigma_B^2/G$. So, the mass contribution
$m_B(\zeta_0,\zeta)$ [eq.(3)] 
of object $B$ at $\zeta$ to an area of $\pi\zeta_0^2$ should
be properly  replaced by $m_B(R_B,\zeta_0,\zeta)$. This gives
a mean surface mass density of objects $B$ around the
object $A$ to be
\begin{equation}
\overline{\Sigma}_{AB}(\zeta_0)=4(1+z_d)^{\epsilon} r_0^2 
		\frac{(\sigma_B^*)^2}{G} 
          F(\zeta_0)
          \int_0^{\infty} \left(\frac{L_B}{L_B^*}\right)^{2/\nu}
          \phi_B(L_B)dL_B,
\end{equation}
where
\begin{equation}
F(\zeta_0)=2\frac{R_B}{\zeta_0}
                   \left(\frac{\zeta_0}{r_0}\right)^{2-\gamma}
		   \int_0^{\zeta_0+R_B}\frac{m(R_B,\zeta_0,\zeta)}{M_B}
	           \chi(\zeta_0,\zeta)d(\zeta/\zeta_0).
\end{equation}
If we introduce the mean spatial mass density of objects $B$,  
$\Omega_B$, in unit of the critical mass density $\rho_0$ 
of the universe, eq.(10) can be written as
\begin{equation}
\overline{\Sigma}_{AB}(\zeta_0)=4\Omega_B\rho_0(1+z_d)^{\epsilon} 
	    r_0f(\zeta_0),
\end{equation}
in which $f(\zeta_0)=(r_0/2R_B)F(\zeta_0)$. This formula was used
by Wu \& Fang (1996) for the estimate of the matter contribution
to  the cluster environments from the cluster-cluster 
correlation function. The advantage of this expression is that
the effects of the observationally determined luminosity function
$\phi_B(L_B)dL_B$ and of the experiential formula eq.(8) are 
represented by a single parameter $\Omega_B$. 

It appears that $F(\zeta_0)$ or $f(\zeta_0)$ is a slowly 
varying function 
of $\zeta_0$ if $\gamma$ is close to $2$. As a result, the 
surface mass density around a given object provided by an 
ensemble of SIS objects that obey a correlation function of 
$\xi(r)\sim r^{-2}$ is nearly a constant, which will be 
numerically verified below for various objects.
The simple reasons are as follows: The total population 
of objects $B$ on the plane perpendicular to the line of 
sight goes as $N_B\sim\zeta$ if their spatial
two-point correlation function has form of $\xi(r)\sim r^{-2}$.
While the projected  mass $m_B$ of a SIS is proportional to
$\zeta$, the expectation of the mass contribution by all the
objects $B$ is then  $m\sim m_BN_B\sim \zeta^2$. Finally, the
expected surface mass density of objects $B$ around $A$, 
$\overline{\Sigma}_{AB}=m/\pi\zeta^2$, remains roughly unchanged. 

We now have two ways to evaluate the surface mass density
around object $A$ contributed by
the two-point correlation function $\xi_{AB}(r)$:
(I)The conservative estimate of 
$\overline{\Sigma}_{AB}$ [eq.(10)],  provided by 
a SIS density profile for $B$ with a proper radius cutoff 
and an observationally determined luminosity function
for the distribution of population $B$;
(II)The optimal estimate of 
$\overline{\Sigma}_{AB}$ [eq.(12)], assuming again 
a truncated SIS density
profile for $B$ but characterizing the distribution of
all the population $B$  by   
their matter density parameter $\Omega_B$ in the universe, 
especially when $\Omega_B=\Omega_0$ where
$\Omega_0$ is the present mean mass density parameter of the universe.
We will employ these two models to 
compute the matter contributions 
from galaxy-galaxy, cluster-cluster, cluster-galaxy and 
quasar-galaxy correlations. The cutoff radii of galaxies ($R_g$)
and clusters of galaxies ($R_c$) are assumed 
to be $R_g\approx0.2\;h^{-1}$ Mpc and $R_c\approx1.5\;h^{-1}$ Mpc, 
respectively, according to the suggestions by 
dynamical analysis and observations (e.g. Bahcall, Lubin \& Dorman 1995). 
In the numerical calculations below
we further require that the minimum separation $r_{min}$ in 
the correlation function $\xi_{AB}(r)$
be greater than the cutoff radius of object $A$.

\subsection{Galaxy-galaxy correlation}

We take the two-point galaxy-galaxy correlation function 
$\xi_{gg}(r)$ from the CfA survey (Davis
\& Peebles, 1983), which provides $r_0=5.4\pm0.3\;h^{-1}$ Mpc and 
$\gamma=1.77\pm0.04$ at scale 10 kpc$\leq r\leq10\;h^{-1}$ Mpc. We 
classify the galaxy population as the early-type galaxies E/S0 and
the spiral galaxies S in terms of their morphologies and adopt the
composition given by Fukugita \& Turner (1991): 
$(\gamma_1,\gamma_2,\gamma_3)=(12\pm2\%,19\pm4\%,69\pm4\%)$,
where $\{i\}=(1,2,3)=$(E,S0,S). The distribution 
of  different galaxies with luminosity is described by the Schechter
luminosity function 
\begin{equation}
\phi_{g,i}(L_g)dL_g=\phi^*_g(L_g/L^*_{g,i})^{\alpha_g}
\exp(-L_g/L^*_{g,i})dL_g/L^*_{g,i},
\end{equation}
where $\phi^*_g=(1.56\pm0.34)\times10^{-2}\;h^{3}$ Mpc$^{-3}$, 
$\alpha_g=-1.07\pm0.05$ and 
$(\sigma^*_{g,1},\sigma^*_{g,2},\sigma^*_{g,3})=(225^{+12}_{-20},
206^{+12}_{-20},144^{+8}_{-13})$ km s$^{-1}$. 
Alternatively, the index in eq.(8) reads $\nu=4$ for E/S0 galaxies 
and $\nu=2.6$ for S galaxies, respectively.

Numerical computation shows that 
$f(\zeta_0)\approx8.6$ 
and varies only $9\%$ over distance 
$\zeta_0=0.005$ -- $0.5$ $h^{-1}$ Mpc. Therefore, the overall
surface mass density 
$\overline{\Sigma}_{gg}(\zeta_0)$ 
from the galaxy-galaxy correlation around a given galaxy  
can be approximately taken to be a constant matter sheet:
\begin{equation}
\overline{\Sigma}_{gg}=\left\{
\begin{array}{ll}
0.0027^{+0.0024}_{-0.0015}\;(1+z_d)^{\epsilon}\;h\; 
(F/0.64) \; {\rm g\;cm}^{-2},& \;\;\; ({\rm I});\\
0.011 \Omega_g\;(1+z_d)^{\epsilon}\;h\; 
(f/8.6) \; {\rm g\;cm}^{-2},& \;\;\; ({\rm II}),
\end{array}\right.
\end{equation}
in which the total error includes the errors in the 
adopted galaxy luminosity
function and  the galaxy-galaxy correlation function.
Combining the above results obtained from the two
models yields 
$\Omega_g\approx0.25$, in good agreement with the claim by
Bahcall et al. (1995).

\subsection{Cluster-cluster correlation}

Cluster-cluster correlation function $\xi_{cc}(r)$ is represented by 
$r_0=(20\pm4.3)\;h^{-1}$
Mpc and $\gamma=1.8$ over scale of $\sim5\;h^{-1}$ Mpc --- $\sim75\;
h^{-1}$ Mpc (Postman, Huchra \& Geller 1992). We employ both the 
X-ray luminosity function and the mass function of galaxy clusters
to estimate the cluster matter contribution from $\xi_{cc}(r)$ 
to a given cluster at $z_d$.

The X-ray luminosity function of clusters of galaxies  shows
a strong evolution with the cosmic epoch. Within the low redshift 
of $z_d\sim0.15$ (Edge et al. 1990)
\begin{equation}
\phi_c(L_x)dL_x=\phi_c^*L_{x,44}^{-\alpha_c}\exp(-L_{x,44}/L_{x,44}^*)
dL_{x,44},
\end{equation}
in which $\phi_c^*=10^{-6.57\pm0.12}\;h_{50}^{3}$ Mpc$^{-3}$, 
$\alpha_c=1.65\pm0.26$,
$L_{x,44}^*=8.1_{-2.3}^{+5.7}$ and $L_{x,44}$ is the X-ray
luminosity in units of $L_{44}=10^{44}\;h_{50}^{-2}$ erg s$^{-1}$. 
For the distant 
clusters of galaxies at redshift ranging from 0.14 to 0.6 
(Henry et al. 1992),
\begin{equation}
\phi_c(L_x)dL_x=\phi_c^*L_{x,44}^{-\alpha_c}dL_{x,44}.
\end{equation}
For the three redshift shells $\{(0.14,0.20),(0.20,0.30),(0.30,0.60)\}$,
$\alpha_c=\{2.19\pm0.21,2.67\pm0.26,3.27\pm0.29\}$ and 
$\phi_c^*=\{5.85\pm0.25,6.82\pm0.51,12.33\pm3.87\}\times10^{-7}
\;h_{50}^{3}$ 
Mpc$^{-3}(L_{44})^{\alpha_c-1}$.  We utilize the {\it quasi-}Faber-Jackson
relation for the X-ray selected galaxy clusters to convert  X-ray
luminosity into velocity dispersion (Quintana \& Melnick 1982;
see also Wu \& Hammer 1993), 
\begin{equation}
L_x=10^{32.72}\sigma_c^{3.94}\;h_{50}^{-2}\;{\rm erg\;s}^{-1}.
\end{equation}
The more recent optical and X-ray observations provide essentially
a similar result (see Wu \& Mao 1995 and references therein):
$L_x=10^{32.64}\sigma_c^4\;h_{50}^{-2}\; {\rm erg\;s}^{-1}$. 
Moreover, we introduce a cluster velocity dispersion 
cutoff $\sigma_{c,min}$ at the faint end of
the luminosity distribution to ensure the convergence of the
integration $\int \sigma_c^2\phi_c(L_x)dL_x$. We choose 
$\sigma_{c,min}=508$ km s$^{-1}$ which corresponds to an $M_c^*$ 
cluster (see below). The resulting 
surface mass density from $\xi_{cc}(r)$ in terms of model I reads
\begin{equation}
\overline{\Sigma}_{cc}=\left\{
\begin{array}{ll}
0.66_{-0.33}^{+0.83}\times10^{-4}\;h\;(F/0.27)\;(1+z_d)^{\epsilon}\;
                     \;{\rm g\;cm}^{-2},\; & \;\;0<z_d<0.15;\\
2.0_{-0.8}^{+1.4}\times10^{-4}\;h\;(F/0.27)\;(1+z_d)^{\epsilon}\;
                     \;{\rm g\;cm}^{-2},\; & \;\;0.14<z_d<0.20;\\
2.8_{-1.0}^{+1.1}\times10^{-4}\;h\;(F/0.27)\;(1+z_d)^{\epsilon}\;
                     \;{\rm g\;cm}^{-2},\; & \;\;0.20<z_d<0.30;\\
8.3_{-4.2}^{+4.2}\times10^{-4}\;h\;(F/0.27)\;(1+z_d)^{\epsilon}\;
                     \;{\rm g\;cm}^{-2},\; & \;\;0.30<z_d<0.60,\\
\end{array} \right.
\end{equation}
in which the variation of 
$f=(r_0/2R_c)F$ is less than $10\%$ from $f=1.8$ for
$\zeta_0=0.5$ -- $10\;h^{-1}$ Mpc.

A relatively simple way of estimating $\overline{\Sigma}_{cc}$ is to use 
the cluster mass function established by Bahcall \& Cen (1993):
\begin{equation}
n_c(>M_c)=\phi_c^*\;(M_c/M_c^*)^{-1}\;\exp(-M_c/M_c^*),
\end{equation}
where $\phi_c^*=4\times10^{-5}\;h^{3}$ Mpc$^{-3}$ 
and $M_c^*=1.8\times10^{14}\;h^{-1}$ $M_{\odot}$. 
$M_c$ refers to the cluster mass within $R_c=1.5$ 
$h^{-1}$ Mpc radius sphere of the cluster center, which is related to
our cluster model through $M_c=2\sigma_c^2R_c/G$. A straightforward
computation gives
\begin{equation}
\overline{\Sigma}_{cc}=\left\{
\begin{array}{ll}
1.3^{+0.6}_{-0.5}\times10^{-4}\;(1+z_d)^{\epsilon}\;h\; 
(F/0.27) \; {\rm g\;cm}^{-2},& \;\;\; ({\rm I});\\
0.0083 \Omega_c\;(1+z_d)^{\epsilon}\;h\; 
(f/1.8) \; {\rm g\;cm}^{-2},& \;\;\; ({\rm II}),
\end{array}\right.
\end{equation}
in which we have utilized the limit $\sigma_c\geq \sigma_{c,min}$.
We have also shown the result of model II, represented by 
the cluster matter density parameter $\Omega_c$ in
the universe, which is consistent with the finding of 
Wu \& Fang (1996). Apparently, 
the result given by model I in the above equation
is compatible with eq.(18) derived from the cluster X-ray 
luminosity function at low redshift $z_d<0.3$.
Note that a comparison of the results by the two models
in eq.(20) yields a relatively small value of the 
cluster matter density: $\Omega_c=0.016$. This is primarily 
due to the adopted low limit 
$\sigma_c\geq\sigma_{c,min}=508$ km s$^{-1}$. 
We will not explore how $\sigma_{c,min}$ affects the estimate of
$\overline{\Sigma}_{cc}$ because the maximum cluster matter 
contribution can be figured out simply by setting $\Omega_c=\Omega_0$
in eq.(20).

\subsection{Cluster-galaxy correlation}

It has been found that the shape of the cross correlation function 
of Abell clusters with  Lick galaxies $\xi_{cg}(r)$  is slightly 
steeper than $\xi_{gg}(r)$ and $\xi_{cc}(r)$, with $\gamma$ ranging
from $1.7$ to $2.5$ (Lilje \& Efstathiou, 1988), while
the amplitude of $\xi_{cg}(r)$ has not been well constrained.  We adopt
the approximate form by Peebles (1993): $r_0=15\pm3\;h^{-1}$
Mpc and $\gamma=2$, which is available on scale roughly 
$0.5\;h^{-1}$ Mpc$<r<40\;h^{-1}$ Mpc. 
We compute both the matter contribution $\overline{\Sigma}_{gc}$
of clusters surrounding a given
galaxy and the matter contribution $\overline{\Sigma}_{cg}$ 
of galaxies
surrounding a given cluster. Replacing $\xi_{gg}$ in eq.(14) and
$\xi_{cc}$ in eq.(20) by $\xi_{cg}$ gives
\begin{equation}
\overline{\Sigma}_{gc}=\left\{
\begin{array}{ll}
2.7^{+1.1}_{-1.0}\times10^{-4}\;(1+z_d)^{\epsilon}\;h\; 
(F/1) \; {\rm g\;cm}^{-2},& \;\;\; ({\rm I});\\
0.017 \Omega_c\;(1+z_d)^{\epsilon}\;h\; 
(f/5) \; {\rm g\;cm}^{-2},& \;\;\; ({\rm II}),
\end{array}\right.
\end{equation}
and 
\begin{equation}
\overline{\Sigma}_{cg}=\left\{
\begin{array}{ll}
0.0031^{+0.0038}_{-0.0013}\;(1+z_d)^{\epsilon}\;h\; 
(F/0.1) \; {\rm g\;cm}^{-2},& \;\;\; ({\rm I});\\
0.013 \Omega_g\;(1+z_d)^{\epsilon}\;h\; 
(f/3.7) \; {\rm g\;cm}^{-2},& \;\;\; ({\rm II}).
\end{array}\right.
\end{equation}
Here $f$ appears to be nearly a constant  over
scale $0.01\;h^{-1}$ Mpc$<\zeta_0<10\;h^{-1}$ Mpc in eq.(21).
However,  because of
the small radius cutoff of  density profile and the $r^{-2}$ 
distribution of spatial number density, galaxies lead to 
a relatively large variation of up to $48\%$ from $f=3.7$ for 
$\zeta$ ranging from 0.5 to $10\;h^{-1}$ Mpc in eq.(22).

\subsection{Quasar-galaxy correlation}

Observations have shown that quasars are preferentially located in 
region of higher than average galaxies, usually in small groups
of galaxies, and behave in a similar way to 
the galaxy-galaxy correlation: $\xi_{qg}(r)\sim \xi_{gg}(r)$ 
 (Bahcall \& Chokshi 1991; references therein). 
This arises from the fact
that both quasars and galaxies trace the same large-scale structure of 
the universe.  The proportional coefficient up to 
$r=0.25\;h^{-1}$ Mpc is roughly $2.3$ for
the optically-selected quasars at $z_d<0.5$,
and 2.8 and 8 for radio-loud quasars at $z_d<0.5$ and 
$\overline{z}_d\sim0.6$, respectively. 
Nevertheless, we assume that $\xi_{qg}(r)$ continues
as $\sim r^{-1.8}$ to a maximum separation of $10\;h^{-1}$ Mpc as for  
the galaxy-galaxy correlation. 
We have checked an alternative model proposed by
Bahcall \& Chokshi (1991) that $\xi_{qg}(r)$ has a steeper shape with 
$\gamma=2.5$ beyond and also normalized at $r=0.25\;h^{-1}$ Mpc.
It is found that these two assumptions have resulted 
in the roughly same value of $f$.
Nevertheless,  the minimum separation
between the quasar and its neighbor galaxies can be in principle 
taken to be $r_{min}\approx0$. This leads to 
$f=39$ with a variation of smaller than $28\%$
over distance $0.01\leq\zeta_0\leq0.1\;h^{-1}$ Mpc. 
Thus, the galactic matter contribution around a quasar can be 
obtained simply by increasing $\overline{\Sigma}_{gg}$
by a constant proportional factor:
\begin{equation}
\overline{\Sigma}_{qg}\approx \left\{
\begin{array}{ll}
2.3\tau\overline{\Sigma}_{gg},
&\;\;\;\; {\rm for\; optical\; quasars\; at} \;z_d<0.5;\\
2.8\tau\overline{\Sigma}_{gg},
&\;\;\;\; {\rm for\; radio\;loud\; quasars\; at} \;z_d<0.5;\\
8\tau\overline{\Sigma}_{gg},
&\;\;\;\; {\rm for\; radio\;loud\; quasars\; at} \;\overline{z}_d\sim0.6,
\end{array} \right.
\end{equation}
where $\overline{\Sigma}_{gg}$ is given in eq.(14) and $\tau$ accounts for
the change of the function $F$ or $f$: $\tau=39/8.6$.
The very recent observations of 20 luminous quasars at $z<0.3$ with
HST (Fisher et al. 1996) have found that the ratio of $\xi_{qg}$ to
$\xi_{gg}$ is $3.8\pm0.8$. This factor is basically
consistent with the result of Bahcall \& Chokshi (1991).


\section{Applications in Statistical Lensing}

\subsection{Gravitational lensing by a uniform matter sheet}

A uniform matter sheet $\overline{\Sigma}$ was introduced by Turner, 
Ostriker \& Gott (1984) to model the cluster matter surrounding a galaxy.
Basically,  gravitational lensing by a massive object having a deflection
angle $\mbox{\boldmath$\alpha$}$ with assistance of a uniform mass sheet
$\overline{\Sigma}$ 
is described by the following lensing equation:
\begin{equation}
\left(1-\frac{\overline{\Sigma}}{\Sigma_{crit}}\right) 
\mbox{\boldmath$\theta$}
-\mbox{\boldmath$\beta$}=\frac{D_{ds}}{D_s}
\mbox{\boldmath$\alpha$},
\end{equation}
where 
\begin{equation}
\Sigma_{crit}=\frac{c^2}{4\pi G}\frac{D_s}{D_{ds}D_d}
\end{equation}
is the critical surface mass density, $\mbox{\boldmath$\beta$}$
and $\mbox{\boldmath$\theta$}$ are the positions of the source and of
the corresponding images, and $D_d$, $D_s$ and $D_{ds}$ are
the angular diameter distances to the lens,
to the source and from the lens to the source, respectively.
It appears that the contribution of a matter sheet to 
gravitational lensing becomes to be significant only when 
$\overline{\Sigma}$ 
reaches a value comparable to $\Sigma_{crit}$.  
$\Sigma_{crit}$ tends to infinity
at both $D_d=0$ ($z_d=0$) and $D_d=D_s$ ($z_d=z_s$). 
For a typical source (e.g quasar) at $z_s=2$ the minimum
value of $\Sigma_{crit}$ is 0.82  $h$ g cm$^{-2}$.
It turns out, according to the calculations 
made in the above section, that
the clustering of galaxies and galaxy clusters
can provide a matter density of as high as
$\overline{\Sigma}/\Sigma_{crit}\sim10^{-2}$ around a
given galaxy or cluster, while the galactic matter around
a given quasar can reach 
$\overline{\Sigma}_{qg}/\Sigma_{crit}\sim10^{-1}$,
provided that $\Omega_g\sim\Omega_c\sim\Omega_0=1$.
This environmental matter may raise the amplitude of gravitational 
lensing associated with foreground galaxies, clusters of galaxies and 
quasars.  In the following subsections 
we explore one of the consequences   
arising from magnification bias due to the presence of
the clustering of galaxies and clusters of galaxies: 
the associations of the angular positions of
background quasars with foreground
galaxies, clusters of galaxies and quasars. 
To do this we will adopt a stable clustering model
$\epsilon=-1.2$, which has been shown to be consistent with the universal 
relation $\xi_{AB}(r)\sim(r/r_0)^{-1.8}$ for 
clustering of galaxies, clusters of galaxies
and quasars (Bahcall \& Chokshi 1991). 
Alternatively, the surface mass
density given by $\xi_{AB}(r)$ refers to the value over a comoving
area. A factor of $(1+z_d)^2$ should be multiplied to get the 
physical surface mass density in an $\Omega_0=1$ universe.

\subsection{Quasar enhancement factor}

Quasar overdensity behind a massive object is usually
characterized by the quasar enhancement factor $q$, which is the
ratio of the observed quasar surface number density over the 
association area around the foreground object
to the intrinsic one $N(<m_b)$ above a limiting magnitude $m_b$. 
In practice, $N(<m_b)$ is obtained using the quasar number density
in the remaining area far away from the foreground object. Moreover, 
the ``unaffected background hypothesis'' is often employed, i.e., the
observed quasar number count is not significantly different from
the intrinsic one $N(<m_b)$. In order to compare with the observations
that search for quasars over an area with distance ranging  
from $\theta_1$ to $\theta_2$ from the foreground object, 
we use the average enhancement factor $\overline{q}$:
\begin{equation}
\overline{q}=\frac{\int_{\theta_1}^{\theta_2}q2\pi\theta d\theta}
                      {\pi(\theta_2^2-\theta_1^2)},
\end{equation}
where (Narayan 1989)
\begin{equation}
q=\frac{N(<m_b+2.5\log \mu)}{N(<m_b)}\;\frac{1}{\mu},
\end{equation}
and $\mu$ is the lensing magnification introduced by all the 
gravitational matter associated with foreground object. Eq.(27)
has accounted for both the magnification bias ($2.5\log\mu$) and the 
area distortion ($1/\mu$) because of light deflection near the
lens. For the optically-selected quasars we  adopt the 
number-magnitude relation from Boyle, Shanks \& Peterson (1988)
\begin{equation}
\begin{array}{ll}
N(<m_b)=4.66\times10^{0.86(m_b-19.15)},\;\;\; & m_b<19.15;\\
N(<m_b)=-10.95+15.61\times10^{0.28(m_b-19.15)},\;\;& m_b>19.15,
\end{array}
\end{equation}
which is valid within $m_b<21$ and $z<2.2$.
If the foreground object has a spherical matter distribution,
the magnification induced by this lens with an additional uniform
mass sheet $\overline{\Sigma}$ becomes
\begin{equation}
\mu=\left|\left(1-\frac{\overline{\Sigma}}{\Sigma_{crit}}\right)^2
     \left(1-\frac{\theta_E}{\theta}
             \frac{\alpha(\theta)}{\alpha(\theta_E)}\right)
     \left(1-\frac{\theta_E}{\alpha(\theta_E)}
              \frac{d\alpha(\theta)}{d\theta}
             \right)
    \right|^{-1},
\end{equation}
where $\theta_E$ is the Einstein radius corresponding to 
$\mbox{\boldmath$\beta$}=0$ in eq.(24).

\subsection{Quasar-galaxy associations}

Statistical evidences for 
quasar-galaxy associations have been accumulated for a decade
since the first observation by Tyson (1986).  
It is generally believed that  statistical lensing by galaxies 
as well as their host clusters is most likely to be the cause.
Here, we investigate how galaxy environments, namely, 
the galaxy-galaxy correlation and the cluster-galaxy correlation, 
affect the lensing properties of quasar-galaxy associations. 
We again utilize a SIS model for the mass density profile of a galaxy
and the same morphological composition $\gamma_i$ ($i=1,2,3$)
for galaxy population as we adopted in section 2.2. 
For the galactic matter (SIS) with a uniform matter sheet 
$\overline{\Sigma}_g\equiv\overline{\Sigma}_{gg}+\overline{\Sigma}_{gc}$ 
provided by $\xi_{gg}(r)$ and $\xi_{cg}(r)$, eq.(29) reduces to
\begin{equation}
\mu=\frac{\theta}{\theta-\theta_{E}}
\frac{1}{(1-\frac{\overline{\Sigma}_{g}}{\Sigma_{crit}})^2},
\end{equation}
where 
$\theta_{E}=\theta_{gE}(1-\overline{\Sigma}_{g}/\Sigma_{crit})^{-1}$ 
and $\theta_{gE}=4\pi(\sigma_g/c)^2(D_{ds}/D_s)$ are the 
the Einstein radius with and without $\overline{\Sigma}_{g}$, respectively.
The expected quasar enhancement factor $\langle q\rangle$
is obtained by averaging $\overline{q}$ over the luminosity
and spatial distributions of galaxies:
\begin{equation}
\langle q\rangle=\frac{\int_0^{z_s}4\pi D_d^2(1+z_d)^3
			(dr_{prop,z_d}/dz_d)dz_d
       \left(\displaystyle\sum_i\int_{L_{min,i}}^{\infty}
       \overline{q}_i\;\gamma_i\phi_i(L_g)dL_g\right)}
                {\int_0^{z_s}4\pi D_d^2(1+z_d)^3
			(dr_{prop,z_d}/dz_d)dz_d
       \left(\displaystyle\sum_i\int_{L_{min,i}}^{\infty}
                       \gamma_i\phi_i(L_g)dL_g\right)},
\end{equation}
in which $dr_{prop,z_d}$ is the proper distance within $dz_d$ of $z_d$,
$L_{min}$ is the luminosity of the faintest galaxies in the sample and
a constant comoving number density of galaxies has been presumed.
In effect, the denominator 
is the total number of galaxies with $L>L_{min}$
and $z<z_s$. Note that the observations actually measure the
surface number density of low-redshift galaxies around 
high-redshift quasars,  giving rise to the so-called 
galaxy enhancement factor (Wu 1994).  
In principle, the foreground
galaxy enhancement factor should be equal to  
the background quasar enhancement factor $q$ in the sense that
one counts only the quasar-galaxy pairs within a given distance 
range in a complete sample. 
A detailed numerical computation of eq.(31) using the 
parameters in  the observations of quasar-galaxy associations 
has been recently made by Wu, et al. (1996). We present in Table 1
the observed and expected quasar enhancement factors behind galaxies
with and without the environmental matter $\overline{\Sigma}_{g}$.
It appears that, as compared with galaxies alone  as lenses,
the mean quasar enhancement
factor $\langle q\rangle$ is typically increased by
$\sim0.01$ and $\sim0.1$ for model I and 
model II (assuming that $\Omega_g=\Omega_c=\Omega_0=1$), respectively.
Therefore, $\langle q\rangle$ is
little affected by the galaxy environments if galaxies
and clusters of galaxies  contribute no more than  $<25\%$ to 
the matter of the universe (model I), 
while the theoretical predictions fit better 
the measurements of the quasar-galaxy associations 
with the help of the environmental matter
if $\Omega_g\sim\Omega_c\sim\Omega_0=1$ (model II).

\placetable{table-1}

\subsection{Quasar-cluster associations}

A significant quasar overdensity has been recently detected on scale of
$\sim10$ arcminutes behind foreground clusters 
(Rodrigues-Williams \& Hogan 1994; Wu \& Han 1995; Rodrigues-Williams 
\& Hawkins 1995; Seitz \& Schneider 1995).
It has become clear today that 
cluster matter alone cannot account for the large quasar enhancement
in the scenario of gravitational lensing, and large-scale matter clumps
traced by clusters of galaxies should be invoked. 
Wu \& Fang (1996) have analyzed 
the four measurements of the quasar-cluster associations and found that
the surface mass density required to produce the reported 
quasar enhancement
factors should be as high as $\sim0.2$ $h$ g cm$^{-2}$. They 
only computed the matter contribution $\overline{\Sigma}_{cc}$
from the cluster-cluster correlation function $\xi_{cc}(r)$ and 
concluded that the resulting $\overline{\Sigma}_{cc}$ is at least an
order of magnitude lower than $\sim0.2$ $h$ g cm$^{-2}$ even if 
the total cluster matter is large enough to close the universe.
This has been clearly shown in eqs.(18) and (20). 
The present paper has further demonstrated  that  
the discrepancy cannot be resolved by involving the 
cluster-galaxy correlation $\xi_{cg}(r)$:
the galaxies in the proximity of clusters of galaxies
following  $\xi_{cg}(r)$
provide a mean surface mass density
$\overline{\Sigma}_{cg}$ [eq.(22)] comparable with 
$\overline{\Sigma}_{cc}$ [eq.(20)], which is insufficient to
explain the associations of the distant quasars at 
$z_s\sim1$ -- $2$ with the nearby clusters at 
$\overline{z}_d\sim0.15$.

\subsection{Quasar-quasar associations}

The presence of close quasar pairs with separation of a few arcseconds 
but with different redshifts has been a puzzle for over two decades.
Four events have been so far detected among $\sim$1000 -- 2000 surveyed
quasars (Burbidge et al. 1996). 
While the probability of finding four quasar pairs 
within $5^{\prime\prime}$ based on accidental configuration turns out to be 
relatively small,  it is argued that quasars probably have significant 
non-cosmological redshift components.  An alternative to this
non standard model is to attribute the quasar-quasar associations to
magnification bias by foreground matter inhomogeneities, as was
suggested by Schneider (Burbidge et al. 1996). However, there have been
no any realistic lensing models which could be constructed to produce
the reported quasar-quasar associations. Now, the quasar-galaxy correlation
function $\xi_{qg}(r)$ may provide a solution.

Quasars are preferentially located in the over-density regions which
have a mean surface mass density $\overline{\Sigma}_{qg}$. 
Thus, the probability
of detecting a background companion quasar near a foreground one within
a given range is a factor of $q$ higher than that over the similar area in
the rest of the sky, where the enhancement factor $q$ is given by eq.(27).
Note that $q$ is independent of the magnitude of foreground quasar,
which goes down to 20.7 in the four examples of the 
quasar-quasar associations. 
All the four background companion quasars in the reported events are 
brighter than $\sim19.15$ in $B$ magnitude, the turn-over point  
in the quasar number-magnitude relation [eq.(28)].
This removes the concern by Schneider (Burbidge et al. 1996) that 
quasar counts are too flat at faint magnitude to provide a large
enhancement factor.   

It is rather difficult today to figure out the statistical significance 
for the reported quasar-quasar associations since they did not result
from a uniform quasar sample with a definite limiting magnitude.
Furthermore, the quasar-galaxy correlation  increases considerably
with cosmic epoch, which has not been well constrained. 
While the foreground quasars in the four known examples
have redshifts ranging from 0.44 to 1.62, we take the relation 
$\xi_{qg}(r)\approx8\xi_{gg}(r)$ established so far 
at the highest redshift $\overline{z}_d\sim0.6$ to estimate 
its contribution to gravitational magnification. 
Recall that the recent HST observations have not found 
the significant difference in the clustering amplitude between the
radio-loud and radio-quiet quasars (Fisher et al. 1996).
Our numerical computation shows that the expected $q$ for 
the four quasar-quasar 
association events ranges from $1.2$ to $1.5$ in model I and
from 2 to 8 in model II,  where
we have utilized the mean value of
$\xi_{gg}(r)$ and given no information about the uncertainties
due to the ignorance of the errors in the adopted $\xi_{qg}(r)$. 
It turns out that about $2$ -- $8$ times more distant quasars would be 
found near the positions of foreground quasars than in the rest of the
sky, provided that $\Omega_g\sim\Omega_0=1$ (model II).
This enhancement factor essentially accounts for the probability of
detecting four quasar companions within $5^{\prime\prime}$ of the
primary quasars if the surveyed sample consists of as large as 2000 
quasars.  Recall that the expected quasar pairs based on the random
distribution is about 0.6 (Burbidge et al. 1996). However,
if $\Omega_g\approx0.25$, as indicated by current observations,  
it is unlikely that the situation of the quasar-quasar associations
would be significantly improved by introducing the lensing effect due to 
the quasar environmental matter.

\section{Discussion and conclusions}

The connection of galaxies, clusters of galaxies and quasars 
with their environmental 
matter is characterized by their spatial two-point correlation function
$\xi(r)$. Assuming a SIS model for matter 
distribution in galaxy and cluster of galaxies, 
we have computed the expected matter contributions of the surrounding
galaxies and galaxy clusters following $\xi(r)$ 
to a given galaxy, cluster or quasar. We have then employed 
these surface mass densities from $\xi(r)$ for the estimate of
the background quasar enhancement factor in the vicinities of 
foreground galaxies, clusters and quasars 
according to gravitational lensing magnification bias.

Significance of the environmental effect in study of gravitational 
lensing is demonstrated by the ratio of the surface mass density 
$\overline{\Sigma}$ 
provided by $\xi(r)$ to a critical mass density $\Sigma_{crit}$. 
It appears that the environmental galaxy/cluster matter surrounding a
galaxy/cluster has 
$\overline{\Sigma}/\Sigma_{crit}<10^{-3}$, showing a
negligible contribution to gravitational lensing, if their
mass densities are no more than $\sim25\%$ of the mass 
density of the universe (model I).
In this circumstance, it is unlikely that 
the reported quasar-galaxy, quasar-cluster and
quasar-quasar associations can be attributed or fully
attributed  to gravitational lensing by the environmental matter.
However, in the extreme case of $\Omega_g\sim
\Omega_0=1$ and/or $\Omega_c\sim\Omega_0=1$, the amplitudes
of the quasar-galaxy and quasar-quasar associations
can be increased
by their neighbor galaxies and clusters of galaxies,
resulting in an expected value  that  
is basically in agreement with the observed quasar overdensity
behind foreground galaxies and quasars.
Nevertheless, in no case can
the lensing magnification by clusters and
their surrounding matter account for
the recently detected quasar-cluster associations
on scale of $\sim10$ arcminutes, 
in consistent with the claim of Wu \& Fang (1996).

It is noted that the two-point correlation function $\xi(r)$
is observationally and statistically established. Therefore, 
our conclusions hold true only for an ensemble
of objects. It is just appropriate to the associations of 
background quasars with foreground galaxies and
clusters of galaxies. However, the quasar-quasar associations 
might not result from a complete sample and thereby,  
our estimate about the background quasar enhancement factor near foreground
quasars should be employed cautiously for the present data. 

It should also be noticed that  
the critical surface mass density $\Sigma_{crit}$ depends 
on the cosmological models. We have only adopted a matter dominated flat
universe in the present paper.  If the cosmological constant 
$\lambda_0$ is nonzero,
as was suggested recently by the measurement of the Hubble constant 
$H_0\sim80$ km s$^{-1}$ Mpc$^{-1}$ (Ostriker \& Steinhardt 1995;
references therein),  gravitational lensing by $\xi(r)$
would be more efficient than it is in a $\lambda_0=0$ universe. 
For instance, the minimum $\Sigma_{crit}$ reduces to $0.57\;h$
g cm$^{-2}$ for $z_s=2$ in an $\Omega_0=0.2$ and $\lambda_0=0.8$
universe.  It deserves to be investigated
how the cosmological constant affects the present calculations.
Moreover, $\Sigma_{crit}$ appears to be smaller in a low density
universe. So, if $\Omega_0\approx\Omega_g=0.25$,
one would expect a larger environmental effect than 
what we have discussed here for a flat cosmological model. 

Our present results may suffer from 
uncertainties in the evolutionary models of the
correlation function with cosmic epoch.
Observations have indicated 
a significant evolution of galaxy-galaxy correlation: 
Both the correlation
length $r_0$ and the index $\epsilon$ of eq.(2) are found 
to change with redshift (Shepherd et al. 1996; reference therein).
Recall that for quasars at $z_s\sim2$ the lensing galaxies are 
most likely to be at $z_d\sim0.5$, where the evolution effect
cannot be neglected. 
Alternatively, the evolution of the 
quasar-galaxy correlation has been seen in eq.(23),
while foreground quasars in the study of 
the quasar-quasar associations
are actually beyond these redshift ranges. Therefore, our 
predicted amplitude of the quasar-quasar associations may have 
large errors. Fortunately,
the quasar-cluster associations are little
affected by the evolution since all the clusters in the 
measurements of the quasar-cluster associations are at 
relatively low redshift $\overline{z}_d\sim0.15$. It may be 
interesting to study the dependence
of the quasar-galaxy, especially quasar-quasar associations on
the matter clustering at moderate/high redshift, which may offer a 
possible way to test the evolution models of the formation
of large-scale structions.

Finally, a thin-lensing hypothesis has been used throughout
this paper, which is apparently not a good approximation for 
large-matter clumps in the universe.
We should thus point out that 
the results obtained from a simple analysis of the two-point 
correlation functions 
provide only a useful estimate of the amplitude of gravitational lensing 
generated by large-scale structures on various scales.
An extensive study with N-body simulations may give a new
sight into the subject.

\acknowledgments

We thank an anonymous referee for  valuable comments and suggestions.
WXP  was supported by the National Science Foundation of China
and a World Laboratory fellowship.

\clearpage

\begin{table}
\caption{Quasar-galaxy associations. \label{table-1}}

\begin{tabular}{ccccc}  
\tableline
\tableline
\multicolumn{2}{c}{observations$^*$} & 
\multicolumn{3}{c}{expectations}\\
\tableline
authors & $q_{\rm obs}$ & $q_{\rm galaxies}$ & $q$ (I)  & 
$q$ (II) \\
\tableline
Crampton & $1.4\pm0.5$        &$1.11_{-0.04}^{+0.02}$ 
                              &$1.12_{-0.04}^{+0.03}$
                              &$1.23_{-0.03}^{+0.03}$\\
Kedziora-Chudczer &  $\sim1$  &$1.01_{-0.00}^{+0.00}$ 
                              &$1.02_{-0.01}^{+0.01}$
                              &$1.11_{-0.00}^{+0.01}$\\
Magain  & $\sim2.8$           &$1.47_{-0.15}^{+0.12}$ 
                              &$1.49_{-0.17}^{+0.11}$
                              &$1.64_{-0.14}^{+0.12}$\\
Thomas & $1.7\pm0.4$          &$1.08_{-0.03}^{+0.02}$ 
                              &$1.09_{-0.03}^{+0.03}$
                              &$1.20_{-0.02}^{+0.02}$\\
Van Drom  & $\sim1.46$        &$1.04_{-0.03}^{+0.01}$ 
                              &$1.05_{-0.02}^{+0.02}$
                              &$1.16_{-0.01}^{+0.01}$\\
Webster & $\sim2$             &$1.06_{-0.02}^{+0.02}$ 
                              &$1.07_{-0.02}^{+0.03}$
                              &$1.18_{-0.02}^{+0.02}$\\
      &  $1.0\pm0.3$          &$1.05_{-0.02}^{+0.01}$ 
                              &$1.06_{-0.02}^{+0.01}$
                              &$1.10_{-0.01}^{+0.01}$\\
Yee   &  $1.0\pm0.2$          &$1.03_{-0.01}^{+0.01}$ 
                              &$1.04_{-0.01}^{+0.01}$
                              &$1.09_{-0.01}^{+0.01}$\\
      &  $0.9\pm0.1$          &$1.02_{-0.00}^{+0.01}$ 
                              &$1.03_{-0.01}^{+0.01}$
                              &$1.08_{-0.00}^{+0.01}$\\
\tableline
\end{tabular}

\tablenotetext{}{
$^*$Data are taken from Narayan (1992) and Wu et al. (1996).}

\end{table}

\end{document}